Inducing an Incipient Terahertz Finite Plasmonic Crystal in Coupled Two Dimensional Plasmonic Cavities

G. C. Dyer[a], G. R. Aizin[b], S. Preu[c], N. Q. Vinh[d], S. J. Allen[d], J. L. Reno[a], and E. A. Shaner[a]

[a]*Sandia National Laboratories, P.O. Box 5800, Albuquerque, New Mexico 87185*
[b]*Kingsborough College, The City University of New York, Brooklyn, New York 11235*
[c]*Chair for Applied Physics, University of Erlangen-Nuremberg, 91058 Erlangen, Germany*
[d]*Institute for Terahertz Science and Technology, UC Santa Barbara, Santa Barbara, California 93106*

We measured a change in the current transport of an antenna-coupled, multi-gate, GaAs/AlGaAs field-effect transistor when terahertz electromagnetic waves irradiated the transistor and attribute the change to bolometric heating of the electrons in the two-dimensional electron channel. The observed terahertz absorption spectrum indicates coherence between plasmons excited under adjacent biased device gates. The experimental results agree quantitatively with a theoretical model we developed that is based on a generalized plasmonic transmission line formalism and describes an evolution of the plasmonic spectrum with increasing electron density modulation from homogeneous to the crystal limit. These results demonstrate an electronically induced and dynamically tunable plasmonic band structure.

The terahertz (THz) band of the electromagnetic spectrum (100 GHz – 10 THz) presents a unique scientific frontier because of its historically difficult to access position between infrared optics and microwave electronics. To approach the THz band from the microwave domain, the frequency range of electronic circuit elements must be extended. To approach the THz band from photonics, both the low photon energy (4 meV at 1 THz) and comparatively long wavelengths in free space (300 um at 1 THz) pose challenges. Plasmons, fundamental electronic excitations that have light-like characteristics yet drive bound charge oscillations, offer an avenue to bridge this "THz gap." Surface plasmons [1] can scale to THz frequencies provided there is a wavevector defined by geometry [2,3] such as a grating or cavity to compensate for the momentum mismatch between free-space radiation and the sub-wavelength plasmon. Two dimensional (2D) THz plasmons in silicon (Si) [4], III-V materials such as GaAs/AlGaAs [5] and GaN/AlGaN [6] heterostructures, and graphene [7,8] similarly are sub-wavelength compared to free-space electromagnetic waves, but have an additional degree of freedom due to the ability to control electron density, and thus the 2D plasma frequency, through a field effect. It is this latter capability that has generated interest in 2D plasmonic devices [9, 10] such as THz detectors [11-14], mixers [15,16], emitters [17,18], and field enhancement structures [19].

In this letter we present the first experimental observation, to our knowledge, of a coherent coupling between localized plasmonic excitations and formation of a *finite* one-dimensional THz plasmonic crystal in a two dimensional electron gas (2DEG). Whereas in prior work a quasi-infinite plasmonic crystal in a weakly modulated 2DEG was demonstrated [20], here a finite plasmonic crystal is induced in a strongly modulated 2D electron system. The THz plasmonic crystal is observed in an antenna-coupled multi-gate GaAs/AlGaAs high electron mobility transistor (HEMT) where a plasmonic cavity (PC) is formed between device terminals and multiple gate biases provide a 2D electron density modulation in the HEMT channel. Such a solid state system represents the 2D plasmonic analogue of an Esaki-Tsu superlattice [21] or photonic crystal [22]. However, semiconductor superlattices and photonic crystals are defined intrinsically by their material composition and periodicity, whereas the amplitude and period of the 2DEG density modulation and therefore properties of the plasmonic band spectrum can be modified *in-situ* by manipulation of multiple gate voltages.

The plasmonic system depicted in Figs. 1(a) and 1(b) is based on a multi-gate HEMT with a 14 μm long by 10 μm wide channel. The HEMT was fabricated from a double quantum well GaAs/AlGaAs heterostructure (Sandia wafer EA1149) with a total electron density of $n_0 = 4.02 \times 10^{11}$ cm$^{-2}$ and a mobility of $6.14 \times 10^5$ cm$^2$/V s at 12 K. The 20 nm thick quantum wells are 386 nm below the surface and spaced 7 nm from each other. There are three gates, the source gate (G1), center gate (G2), and drain gate (G3), each designed to be 2 μm wide and separated by 2 μm. The fabricated gate widths and separations differ by approximately 10% from the intended dimensions. The source terminal (S) is referenced to ground potential. The G3 was biased beyond threshold, depleting the electrons underneath it and effectively terminating the two dimensional electron gas (2DEG) at its edges. This forms a 10 μm PC between S and G3 where G1 and G2 can tune the electron density beneath them and hence the resonant frequency of the plasma mode.

The gates of the HEMT were biased in a manner similar to previous work [23] with G3 biased beyond threshold ($V_{TH} \sim -2.5\,V$) and an applied drain current of $I_D = +500\,nA$. Under these conditions, incident THz radiation heats the 2DEG in the 10 μm PC between S and G3, resulting in a bolometric electrical response proportional to the absorbed power. With the 2D electron channel pinched off underneath the G3, a barrier with a highly non-linear and temperature dependent current-voltage characteristic is induced. When the device is illuminated, the absorbed radiation reduces the barrier impedance resulting in a change in voltage between the source and drain, $\delta V_{SD} = I_D \delta R_{barrier}$ [23].

We first consider the response spectra of the PC HEMT measured at constant excitation frequency while independently tuning the electron densities $n_1$ and $n_2$ with the respective gates G1 and G2. The normalized electron densities below either gate are given by $n_{(1,2)}/n_0 = (V_{TH} - V_{(G1,G2)})/V_{TH}$. The regions located beneath the biased G1 and G2 form 2 μm long sub-cavities embedded in the 10 μm PC; tuning of $n_1$ and $n_2$ probes the interaction between these sub-cavities. The experimental bolometric response spectra at 370 GHz and 405 GHz with T = 11.5 K are shown in Figs. 2(a) and 2(b), respectively. The plasmonic spectrum displays an evolution from behavior indicative of coherent interaction between multiple sub-cavities to features resembling a single cavity. Where the modulation in electron density between the gated and ungated regions is strong ($n_{(1,2)}/n_0 \lesssim 0.5$), a multitude of anti-crossings between high order plasmon modes associated with the G1 (vertical bands of high intensity) and G2 (horizontal bands of high intensity) sub-cavities develops. The coherence of the plasmon between S and G3 results in a dense spectrum of repelled crossings between adjacent eigenmodes weakly split in energy. Where $n_{(1,2)}/n_0 \gtrsim 0.5$ the resonances correspond to smooth contours of high intensity. This indicates that the 10 µm length from S to G3 and the PC boundary conditions primarily define the resonance condition.

Having established coherence between the plasmonic sub-cavities below G1 and G2, we next examine the bolometric responsivity spectrum at 11.5 K for excitation frequencies from 195 to 450 GHz while identically tuning the gates G1 and G2 such that $n_1 = n_2 \equiv n$. The spectrum shown in Fig. 3(a) is complicated slightly by Fabry-Perot interference in the substrate and cryostat window seen as horizontal bands in the signal. A collection of absorption resonances are observed which shift to lower frequency and increase in density as the electron density $n$ below G1 and G2 is decreased. This behavior is typical of a plasmonic absorption spectrum, where the plasma resonant frequency scales as $\omega_p \propto \sqrt{n}$ [24]. At $n/n_0 = 1$ these plasmon modes correspond to excitations in the PC between S and G3 with quantized wavevectors determined by the PC length of 10 µm and PC boundary conditions. The modes are nearly evenly spaced suggesting that the plasmon is strongly screened by the gates and adjacent metallization because in the screened limit $\omega_p \propto q$ [24] and an equidistant plasmon spectrum in the PC is expected.

However, as the normalized electron density is swept towards $n/n_0 \to 0$ the modes in Fig. 3(a) appear to merge in pairs. The modes that begin near 225 and 275 GHz at $n/n_0 = 1$ form the lowest energy pair of modes. The next set of modes is evident at 350 and 425 GHz at $n/n_0 = 1$. In both cases the higher frequency mode tapers off as $n/n_0 \to 0$, which we interpret as a merging of the high frequency mode with the low frequency mode in each pair. The third pair of eigenmodes is not strongly excited experimentally while still higher order modes are difficult to resolve. This pairing of modes is indicative of a plasmonic band structure forming in this system. In a strongly modulated regime, low frequency plasmons should be mostly localized in the regions with low electron density, the sub-cavities below G1 and G2 [19]. Thus, the mode pairing may be viewed as the energetic splitting of the degenerate eigenmodes in these sub-cavities due to coherent coupling between them.

As the basis for modeling the experimentally observed coherent PC phenomena we take a transmission line (TL) description for the device shown in Fig. 1. A PC consisting of alternating gated and ungated regions of 2DEG, each $L = 2\ \mu m$ long and $W = 10\ \mu m$ wide, is formed between the Ohmic source contact and depleted region below G3. The five regions of 2DEG shown in Fig. 1(a) are treated as discrete segments in a two-port TL network. We assume that the voltage at the Ohmic source contact on one side of the cavity goes to zero. This is equivalent to requiring the plasmon charge density fluctuation vanishes at this boundary. In the equivalent circuit shown in Fig. 1(c) the antenna is represented as a voltage generator $V_{THz}$ in series with source impedance determined by its radiation resistance $R_{ANT}$. It is assumed that for the log-periodic antenna on a GaAs substrate in Fig. 1(d) that $V_{THz}$ and $R_{ANT} \cong 72\ \Omega$ are divided equally between two isolated circuits separated by the barrier induced under the drain gate.

In the TL formalism, a 2DEG of intrinsic density $n$ is described in the Drude model by a distributed impedance $Z_{2DEG} = R_{2DEG} + i\omega L_{2DEG}$ where $R_{2DEG} = m^*/e^2 n\tau W$ and $L_{2DEG} = \tau R_{2DEG}$ are the resistance and kinetic inductance of 2DEG per unit channel length respectively, $\tau$ is the electron momentum relaxation time, $m^*$ is the electron effective mass, and $W$ is the width of the 2DEG channel [25,26]. To describe the experiment we assume that a 2DEG is embedded into a medium with dielectric permittivity $\epsilon$ and occupies the plane $z = 0$. An ideal metal gate is positioned in the plane $z = d$. In the quasi-static approximation the dispersion law for 2D plasmons propagating in this system is determined as $\omega(\omega - i/\tau) = e^2 nq/m^*\epsilon(1 + \coth[qd])$, where $q = q' + iq''$ is the complex plasmon wave vector [24]. The electric potential $\Phi_\omega(x, z)$ of the plasma wave of frequency $\omega$ is connected with the plasma wave charge density per unit length $\rho_\omega(x)$ by the equation $\rho_\omega(x) = C(\omega)\Phi_\omega(x, z = 0)$ where $C(\omega) = W\epsilon q(1 + \coth[qd])$ and $q$ is determined by dispersion relation. The coefficient $C(\omega)$ can be interpreted as electrostatic capacitance per unit length of the equivalent TL. In the fully screened limit ($qd \to 0$), $C(\omega)$ reduces to the standard gate capacitance of a HEMT, $C = W\epsilon/d$. In the unscreened limit ($qd \to \infty$) we recover the expression $C = 2W\epsilon q$ [27]. In a lossy TL ($q'' \neq 0$) the imaginary part of the capacitance represents a shunt conductance in the TL equivalent circuit. With these definitions of the distributed TL elements and the characteristic impedance $Z_0^{-1} = \sqrt{i\omega C(\omega)/(R_{2DEG} + i\omega L_{2DEG})} = \epsilon\omega W(1 + \coth[qd])$, solution of the corresponding telegrapher's equations [28] reproduces the plasmon dispersion law, $q = -i\sqrt{i\omega C(\omega)(R_{2DEG} + i\omega L_{2DEG})}$, as well as spatial distributions of the in-plane electric potential, current and charge density of the plasma wave.

The electric field of the 2D plasmon has a longitudinal component. Therefore, the power carried by the 2D plasma wave cannot be found by direct application of the TL equations [27,29]. The time-averaged complex power carried by the plasma wave can be evaluated using the Poynting vector to yield $P_\omega(x) = \frac{1}{2}\Phi_\omega(x, 0)I_\omega^*(x)\xi^*(\omega, d)$ where $I_\omega$ is the plasmonic current and the form-factor $\xi(\omega, d)$ is determined as

$$\xi(\omega, d) = 1 - \frac{q\left(1-e^{-2q'd}\cos[2q''d]\right)}{2q'\left(1-e^{-2q^*d}\right)} + \frac{q\,e^{-2q'd}\sin[2q''d]}{2q''\left(1-e^{-2q^*d}\right)}. \tag{1}$$

The power transfer can be reduced to the standard TL expression $\frac{1}{2}\tilde{V}(x)\tilde{I}^*(x)$ by defining effective voltage $\tilde{V}(x)$ and current $\tilde{I}(x)$ [29]. We define them as $\tilde{V}(x) \equiv \Phi_\omega(x, 0)$ and $\tilde{I}(x) \equiv \xi(\omega, d)I_\omega(x)$. This definition of effective current is equivalent to redefining the characteristic impedance of the TL as $\tilde{Z}_0 = Z_0/\xi(\omega, d)$. In the fully screened limit, the 2D plasma wave becomes purely transverse and $\xi(\omega, d) = 1$. In the unscreened limit with $q'' = 0$ we obtain $\xi(\omega, d) = 1/2$ [27]. We assume continuity of the voltage $\tilde{V}(x)$ and the power flow $P_\omega(x)$ at the boundaries between segments. This results in the discontinuity of the plasma current $I_\omega$ accounting for the edge charge accumulation at interfaces between the segments as confirmed by direct numerical solution of Maxwell's equations in the 2DEG with step-like changes of the equilibrium electron density and gate screening [30].

Using a transfer matrix method for a segmented TL [28] with the above plasmonic TL definitions, we calculated power absorption in the inhomogeneous PC at different excitation frequencies as a function of the density modulations $n_{(1,2)}/n_0$ under the gates G1 and G2. In this calculation we assumed that the PC between S and G3 is screened uniformly by the gates and surrounding metallization by taking $d = 386\,nm$, the gate-channel separation, as constant. In evaluating the TL effective resistance responsible for the plasmon resonance linewidth we added radiation damping $\tau_{rad}^{-1} = e^2 n / 2m^* \sqrt{\mu_0/\epsilon}$ to the mobility-related damping [31]. The analytical results are shown in Figs. 2(c) and 2(d) for frequencies of 370 GHz and 405 GHz respectively. Here plasmonic resonances correspond to maxima in power absorption because the cavity impedance is always larger than $R_{ANT}$. Discrepancies in the width, position and relative amplitudes of the plasma resonances are most likely due to uncertainty in the grating duty cycle and the fidelity of the modulation boundary, with additional broadening of the resonances due to edge scattering [32]. When density modulation under both gates is strong $(n_{(1,2)}/n_0 \ll 1)$, plasmon modes of different orders localized under G1 and G2 sequentially resonate with each other resulting in the multiple anti-crossing features in Figs. 2(c) and 2(d). These results validate our interpretation of the behavior observed experimentally as shown in Fig. 2(a) and 2(b).

These conclusions are further confirmed by the model results shown in Fig. 3(b) which are in very good quantitative agreement with experiment. These analytical results show a continuous transition from a set of discrete PC modes to the pairing of modes into plasmonic crystal bands as density modulation $n/n_0$ sweeps from 1 to 0. In the strongly modulated limit and in the measured frequency range plasmons are mostly localized under the gates [19]. Interaction between these plasmons lifts degeneracy of the plasmonic spectrum and results in the formation of plasmonic energy bands in a manner similar to the tight-binding approximation in electron band theory.

In order to develop further insight into the formation of a finite plasmonic crystal we calculated the spatial distribution of the electric potential in the plasma wave for two separate bias conditions: (a) $(n/n_0 = 1)$ and (b) $(n/n_0 = .3)$ under both gates. This is shown in Figs. 4(a) and 4(b), respectively. With no density modulation, the spatial voltage distribution in the different plasma modes corresponds to plasmons with quantized wave vectors determined by the finite cavity length and the mode index with fundamental plasma cavity mode at 100 GHz, as illustrated in Fig. 4(a). From Fig. 4(a) we conclude that the experimental modes nearest to $n_{(1,2)}/n_0 = 1$ in both Figs. 2(a) and 2(b) are the 4$^{th}$ harmonic of the unmodulated cavity and in Fig. 3(a) the 2$^{nd}$ and higher order harmonics are observed. With strong density modulation, the modes pair into what is the beginning of the band structure, though the plasmon is still well-confined within the gated region, as illustrated in Fig. 4(b). The fundamental cavity mode near 80 GHz is not paired with another mode. The eigenmode near 460 GHz is associated with the fundamental resonance in the ungated regions of 2DEG and does not pair with a second mode. In fact the ungated regions represent a second plasmonic sub-lattice, but due to its expected resonant frequency above the measured range and the lack of gate tuning it has not been probed in the present experimental system.

The TL formalism developed in this letter is broadly applicable to any 2D plasmonic system including graphene [33,34] provided the appropriate 2D plasmon dispersion law resulting from linear energy-momentum carrier dispersion is employed [35]. Though a TL approach has been suggested for graphene [27], the effects of screening or inhomogeneity due to metal terminals were not included. To consider graphene HEMTs [36,37] as 2D plasmonic devices, especially designs featuring a top gate or having step-like variation of electron density or screening along the channel, a generalized TL approach is required. The high mobility of suspended graphene at room temperature makes it a particularly promising material system for developing components of an integrated THz plasmonic circuit. In contrast, III-V based plasmonic devices operating below 1 THz require cryogenic temperatures (< 100 K) to operate due to the degradation of the carrier mobility-limited plasmon Q-factor at higher temperatures.

In conclusion, we have experimentally demonstrated coupling between plasmonic cavities electronically induced in a homogeneous 2DEG and formation of a THz finite one-dimensional plasmonic crystal. To describe this effect we developed a TL model which is in excellent quantitative agreement with experiment. Given the versatility of the TL approach, such a model could be applied to a multitude of 2D plasmonic systems and devices.

The authors would like to thank J. Mikalopas for help with numerical simulations, M. S. Sherwin, A. C. Gossard, M. C. Wanke, S. Scott and C. D. Nordquist for beneficial discussion and support, and D. Bethke for providing SEM images. This work was performed, in part, at the Center for Integrated Nanotechnologies, a U.S. Department of Energy, Office of Basic Energy Sciences user facility. Sandia National Laboratories is a multi-program laboratory managed and operated by Sandia Corporation, a wholly owned subsidiary of Lockheed Martin Corporation, for the U.S. Department of Energy's National Nuclear Security Administration under contract DE-AC04-94AL85000. The work at Sandia National Laboratories was supported by the DOE Office of Basic Energy Sciences. This material is based upon work supported by the US Air Force Office of Scientific Research, Arlington, VA under Contract Number FA9550-10-C-0172 and Physical Sciences Inc. Andover, MA under agreement number FI011090528. Any opinions, findings and conclusions or recommendations expressed in this material are those of the author(s) and do not necessarily reflect the views of the US AFOSR or Physical Sciences Inc. S. P. acknowledges the Humboldt foundation and the NSF MRSEC program DMR-0520415 (MRL-UCSB) for funding.

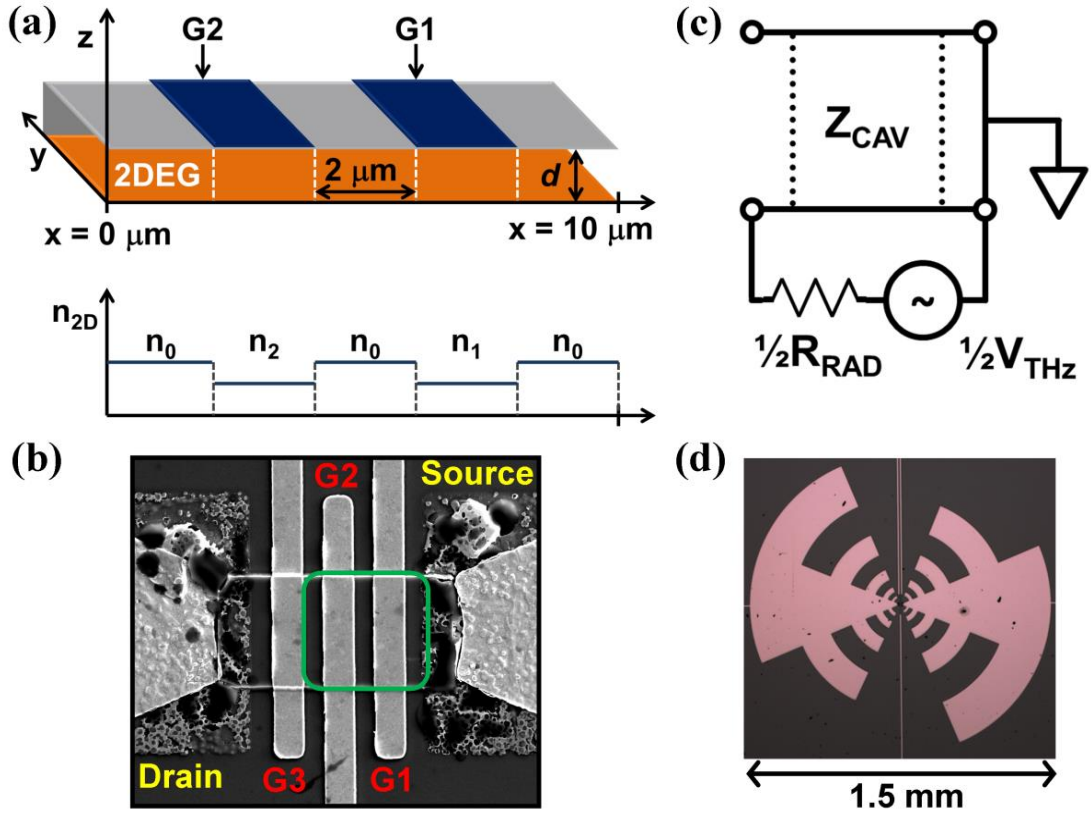

**Figure 1**. (a) Diagram of the plasmonic cavity and coordinate system used for model calculations. Carrier densities in the five regions are indicated schematically. (b) SEM micrograph of the 10 μm by 10 μm plasmonic cavity (highlighted in green) at antenna vertex. Source, drain and center gates are labeled G1, G2 and G3 respectively. (c) Equivalent circuit for the plasmonic cavity where half of the antenna serves as a voltage generator. (d) Log periodic antenna coupling radiation to the plasmonic cavity.

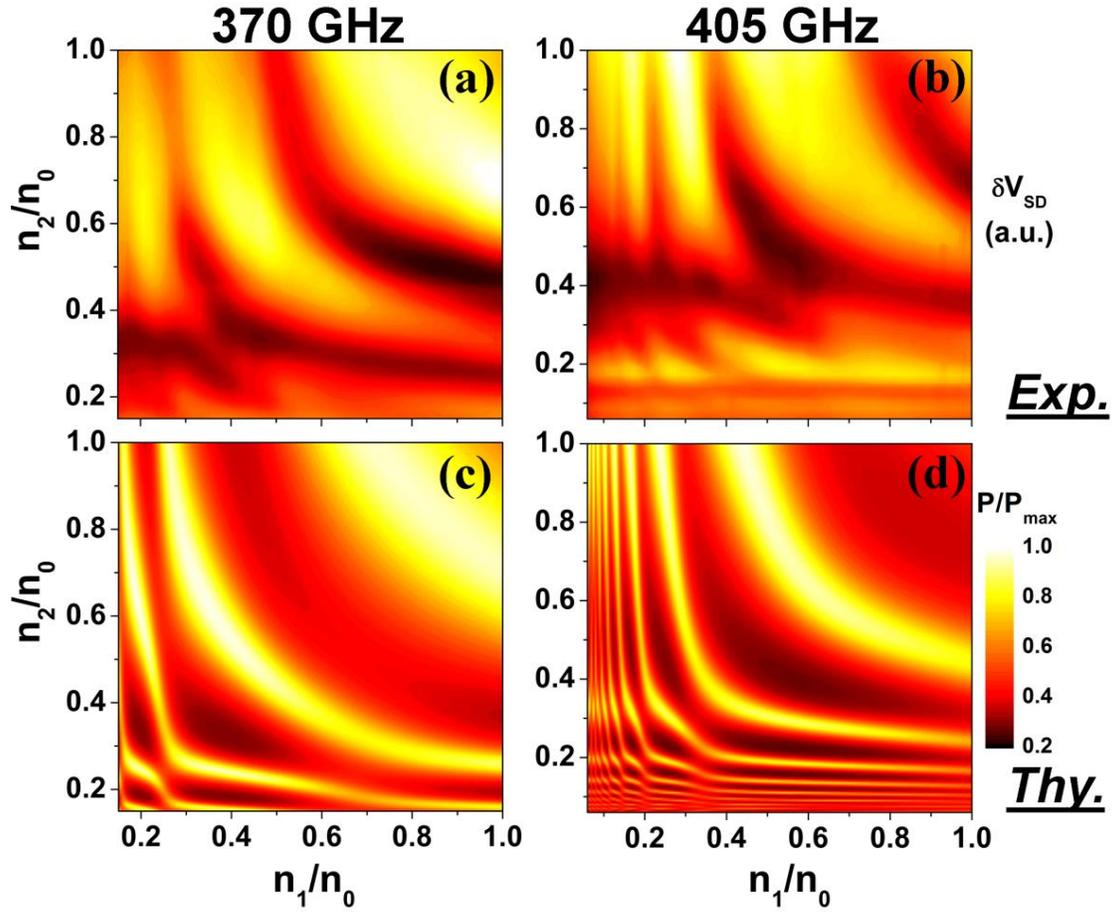

**Figure 2**. Plasmonic cavity THz photoresponse as a function of relative electron densities under G1 ($n_1/n_0$) and G2 ($n_2/n_0$) a with G3 pinched off and T = 11.5 K with (a) 370 GHz and (b) 405 GHz excitation. Model power absorption spectrum as a function of $n_1/n_0$ and $n_2/n_0$ at (c) 370 GHz and (d) 405 GHz for a 10 μm long by 10 μm wide cavity as illustrated in Fig. 1.

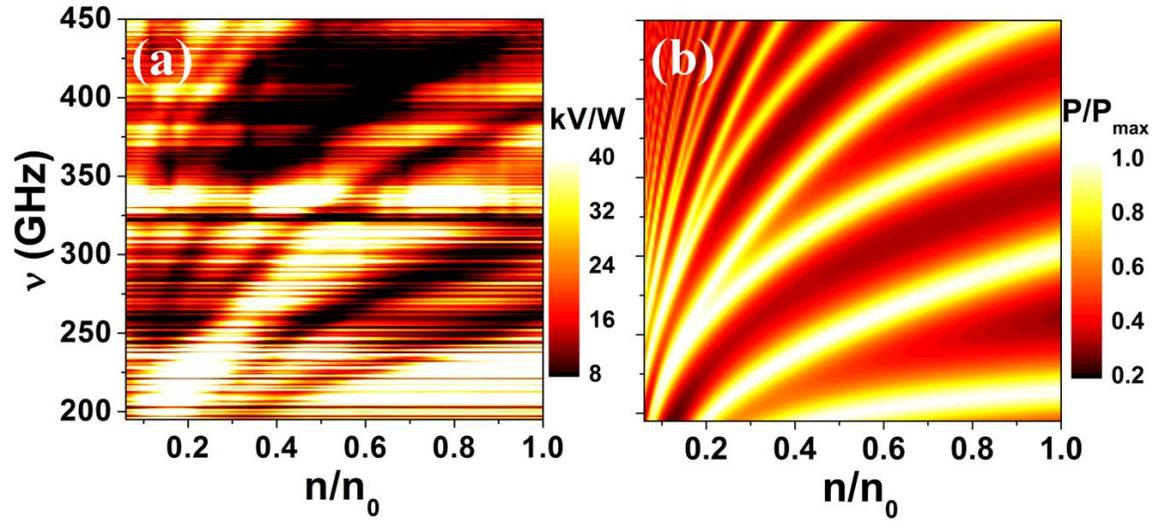

**Figure 3**. (a) Plasmonic cavity responsivity spectrum as a function of frequency and relative electron density $n/n_0$ under the gates with G3 pinched off and G1 and G2 tuned identically at 11.5 K. (b) Model power absorption spectrum for a 10 μm long by 10 μm wide cavity as illustrated in Fig. 1.

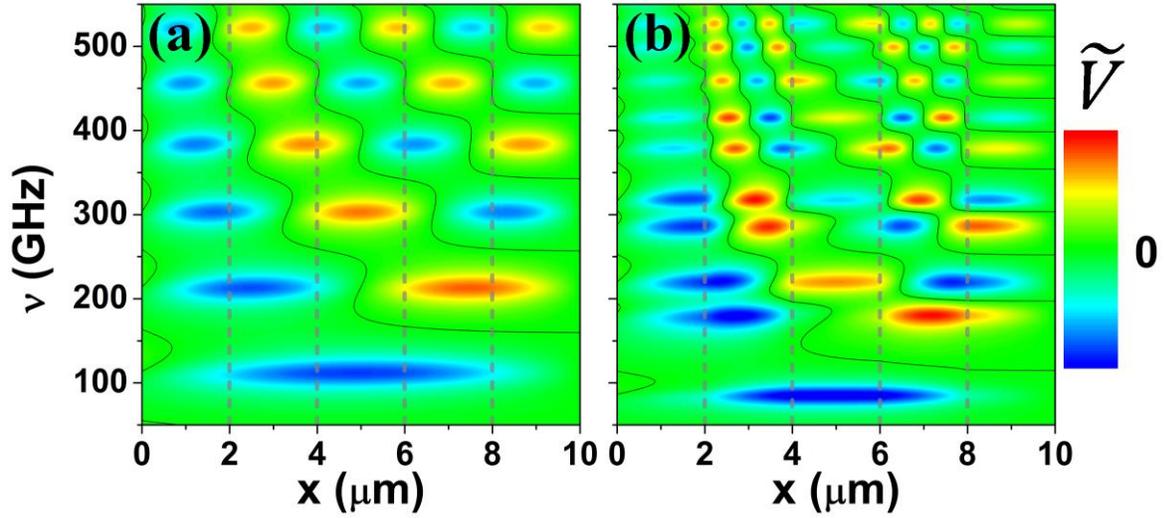

**Figure 4**. The map of the calculated in-plane cavity voltage distribution $\tilde{V}$ as a function of frequency and distance from G3. Dashed lines highlight the boundaries between ungated (0-2, 4-6, 8-10 μm) and gated (2-4, 6-8 μm) regions. The black contours indicate nodes in the voltage distribution. (a) $\tilde{V}$ at $n/n_0 = 1$. (b) $\tilde{V}$ at $n/n_0 = .3$.

14